\documentclass[12pt,draftcls,onecolumn] {IEEEtran}
\usepackage[cmex10]{amsmath}
\usepackage{pifont}
\usepackage{stmaryrd}
\usepackage{mathrsfs}
\usepackage{amsmath}
\usepackage{amssymb}
\usepackage{multirow}
\usepackage{graphicx}
\usepackage{subfigure}
\usepackage{color}
\usepackage[colorlinks,
            linkcolor=black,
            anchorcolor=black,
            citecolor=black
            ]{hyperref}

\newcommand{\bs}{{\mbox{\boldmath{$s$}}}}
\newcommand{\bu}{{\mbox{\boldmath{$u$}}}}

\newcommand{\bw}{{\mbox{\boldmath{$w$}}}}

\newcommand{\bD}{{\mbox{\boldmath{$D$}}}}

\newcommand{\bH}{{\mbox{\boldmath{$H$}}}}

\newcommand{\bS}{{\mbox{\boldmath{$S$}}}}

\newcommand{\bgamma}{{\mbox{\boldmath{$\gamma$}}}}

\begin{document}
\title{OFDM Synthetic Aperture Radar Imaging with Sufficient Cyclic Prefix}
\author{Tian-Xian Zhang and Xiang-Gen Xia, {\em Fellow, IEEE}
\thanks{Tian-Xian Zhang is with the School of Electronic Engineering, University
of Electronic Science and Technology of China, Chengdu, Sichuan, P.R. China,
611731. Fax: +86-028-61830064, Tel: +86-028-61830768, E-mail:
tianxian.zhang@gmail.com. His research was supported by the Fundamental
Research Funds for the Central Universities under Grant ZYGX2012YB008 and by
the China Scholarship Council (CSC) and was done when he was visiting the
University of Delaware, Newark, DE 19716, USA. Xiang-Gen Xia is with the
Department of Electrical and Computer Engineering, University of Delaware,
Newark, DE 19716, USA. Email: xxia@ee.udel.edu. Xia's research was partially
supported by the Air Force Office of Scientific Research (AFOSR) under Grant
FA9550-12-1-0055. }}

\maketitle  

\begin{abstract}
The existing linear frequency modulated (LFM) (or step frequency) and random
noise synthetic aperture radar (SAR) systems may correspond to the frequency
hopping (FH) and direct sequence (DS) spread spectrum systems in the past
second and third generation wireless communications. Similar to the current and
future wireless communications generations, in this paper, we propose OFDM SAR
imaging, where a sufficient cyclic prefix (CP) is added to each OFDM pulse. The
sufficient CP insertion converts an inter-symbol interference (ISI) channel
from multipaths into multiple ISI-free subchannels as the key in a wireless
communications system, and analogously, it provides an inter-range-cell
interference (IRCI) free (high range resolution) SAR image in a SAR system. The
sufficient CP insertion along with our newly proposed SAR imaging algorithm
particularly for the OFDM signals also differentiates this paper from all the
existing studies in the literature on OFDM radar signal processing. Simulation
results are presented to illustrate the high range resolution performance of
our proposed CP based OFDM SAR imaging algorithm.

\end{abstract}


\maketitle

\clearpage
\section{Introduction}\label{Introduction}
Synthetic aperture radar (SAR) can perform well to image under almost all
weather conditions \cite{currie1989radar}, which, in the past decades, has
received considerable attention. Several types of SAR systems using different
transmitted signals have been well developed and analyzed, such as the linear
frequency modulated (LFM) chirp radar \cite{Soumekh1999Synthetic},
linear/random step frequency radar \cite{currie1989radar,
AxelssonGRSRandomStep2007890}, and random noise radar
\cite{XiaoJianAESFOPENSAR20011287, NarayananGRSRandomNoiseSAR20022543,
GuoSuiAESRandomSAR2003489}.

Recently, orthogonal frequency-division multiplexing (OFDM) signals have been
used in radar applications, which may provide opportunities to achieve
ultrawideband (UWB) radar. OFDM radar signal processing was first presented in
\cite{LevanonIEEPMultiF2000276} and was also studied in
\cite{FrankenDoppler2006108,GarmatyukConceptual20081,GarmatyukWidebandOFDM20091,
SturmAnovelOFDM20091,SitOFDMJoint201169, SturmIEEEPWaveformDesign20111236}.
Adaptive OFDM radar was investigated for moving target detection and
low-grazing angle target tracking in
\cite{SenSPLOFDMDetection2009592,SenTSPOFDMWAF2010928,SenTSPOFDMMIMO20103152}.
Using OFDM signals for SAR applications was proposed in
\cite{RicheOFDMSAR12278, RicheOFDMSAR122156,
RicheInvestigationsOnOFDMIEEETGRS20144194,
JungHyoKimNovelOFDMChirpIEEEGRSL2013568, GarmatyukOFDMSAR06237,
GarmatyukGRSOFDMSAR20113780, GarmatyukGRSLOFDMSAR2012808}. In
\cite{RicheOFDMSAR12278,RicheOFDMSAR122156,RicheInvestigationsOnOFDMIEEETGRS20144194},
adaptive OFDM signal design was studied for range ambiguity suppression in SAR
imaging. The reconstruction of the cross-range profiles is studied in
\cite{GarmatyukGRSOFDMSAR20113780, GarmatyukGRSLOFDMSAR2012808}. Signal
processing of a passive OFDM radar using digital audio broadcast (DAB), digital
video broadcast (DVB), Wireless Fidelity (WiFi) or worldwide inoperability for
microwave access (WiMAX) signals for target detection and SAR imaging was
investigated in \cite{BergerSTSPOFDMPassive2010226,ColoneAESOFDMPassive2011240,
GutierrezAESOFDMPassiveSAR2013945,FalconeExperimental2010516,
ColoneAmbiguityFunction2010689,ChettyPassiveWiMAX2010188,QingWiMAX20091}.
However, all the existing OFDM radar (including SAR) signal processing is on
radar waveform designs with ambiguity function analyses to mitigate the
interferences between range/cross-range cells using multicarrier signals
similar to the conventional waveform designs and the radar receivers, such as
SAR imaging algorithms, are basically not changed. The most important feature
of OFDM signals in communications systems, namely, converting an
intersymbol-interference (ISI) channel to multiple ISI-free subchannels, when a
sufficient cyclic prefix (CP) is inserted, has not been utilized so far in the
literature. In this paper, we will fully take this feature of the OFDM signals
into account to propose OFDM SAR imaging where a sufficient CP is added to each
OFDM pulse, as the next generation high range resolution SAR imaging. In our
proposed SAR imaging algorithm, not only the transmission side but also the
receive side are different from the existing SAR imaging methods. To further
explain it, let us briefly overview some of the key signalings in SAR imaging.

To achieve long distance imaging, a pulse with long enough time duration is
used to carry enough transmit energy \cite{skolnik2001Introduction}. The
received pulses from different scatterers are overlapped with each other and
cause energy interferences between these scatterers. To mitigate the impact of
the energy interferences and achieve high resolution, the transmitted pulse is
coded using frequency or phase modulation (i.e., LFM signal and step frequency
signal) or random noise type signals in random noise radar to achieve a
bandwidth $B$ which is large compared to that of an uncoded pulse with the same
time duration \cite{skolnik2001Introduction}. This is similar to the spread
spectrum technique in communications systems. Then, pulse compression
techniques are applied at the receiver to yield a narrow compressed pulse
response. Thus, the reflected energies from different range cells can be
distinguished \cite{skolnik2001Introduction}. However, the energy interferences
between different range cells, that we regard as inter-range-cell interference
(IRCI), still exist because of the sidelobes of the ambiguity function of the
transmitted signal, whose sidelobe magnitude is roughly $\sqrt{N}$ if the
mainlobe of the modulated signal pulse is $N$ \cite{XiaTSPDCFT20003122}. This
IRCI is much significant in SAR imaging \cite{Soumekh1999Synthetic}. Despite of
the IRCI, the existing well-known/used SAR are LFM (or step frequency) SAR
\cite{Soumekh1999Synthetic, AxelssonGRSRandomStep2007890} and random noise SAR
\cite{XiaoJianAESFOPENSAR20011287, NarayananGRSRandomNoiseSAR20022543,
GuoSuiAESRandomSAR2003489}.

We are adopting the OFDM technique \cite{prasad2004ofdm} that is the key
technology in the latest wireless communications standards, such as long term
evolution (LTE) \cite{dahlman20114G} and WiFi. If we think about only one user
in a wireless communications system, there are similarity and difference
between wireless communications systems and SAR imaging systems. The similarity
is that both systems are transmitting and receiving signals reflected from
various scatterers and the numbers of multipaths depend on the transmitted
signal bandwidths. The difference is that in communications systems, the
receiver cares about the transmitted signal, while in SAR imaging systems, the
receiver cares about the scatterers within different range cells (with
different time delays) that reflect and cause multipath signals. The multipaths
cause ISI in communications, while the multipaths cause the IRCI in SAR imaging
systems. The higher the bandwidth is, the more multipaths there are in
communications systems, and the more range cells there are for a fixed imaging
scene (or swath width), i.e., the higher the range resolution is, in SAR
imaging systems. The existing well-known/used LFM (or step frequency) and
random noise SAR systems, in fact, correspond to the two spread spectrum
systems, i.e., frequency hopping (FH) and direct sequence (DS) in CDMA systems
\cite{goldsmith2005wireless}, which work well in non-high bandwidth wireless
communications systems, such as the second and the third generations of
cellular communications, of less than $10$ MHz (roughly) bandwidth. They,
however, may not work well for a system of a much higher signal bandwidth, such
as $20$ MHz in LTE, due to the severe ISI caused by too many multipaths. In
contrast, since OFDM with a sufficient CP can convert an ISI channel to
multiple ISI-free subchannels, as mentioned earlier, it is used in the latest
LTE and WiFi standards. As an analogy, one expects that LFM and random noise
SAR may not work too well for high bandwidth radar systems where there are too
many range cells in one cross range, which cause severe IRCI due to the
significant sidelobes of the ambiguity functions of the LFM and random signals.
On the other hand, in order to have a high range resolution, a high bandwidth
is necessary. Therefore, borrowing from wireless communications, in this paper
we propose to use OFDM signals with sufficient CP to deal with the IRCI problem
as a next generation SAR imaging to produce a high range resolution SAR image.
We show that, in our proposed OFDM SAR imaging with a sufficient CP, the
sidelobes are ideally zero, and for any range cell, there will be no IRCI from
other range cells in one cross range.

Another difference for OFDM communications systems and CP based OFDM SAR
systems is as follows. It is known in communications that, for an OFDM system,
a Doppler frequency shift is not desired, while the azimuth domain (or cross
range direction) in a SAR imaging system is, however, generated from the
relative Doppler frequency shifts between the radar platform and the
scatterers. One might ask how the OFDM signals are used to form a SAR image.
This question is not difficult to answer. The range distance between radar
platform and image scene is known and the radar platform moving velocity is
known too. Thus, the Doppler shifts are also known, which can be used to
generate the cross ranges similar to other SAR imaging techniques, and can be
also used to compensate the Doppler shift inside one cross range and correct
the range migration.

This paper is organized as follows. In Section \ref{System Model}, we propose
our CP based OFDM SAR imaging algorithm. In Section \ref{Simulations}, we
present some simulations to illustrate the high range resolution property of
the proposed CP based OFDM SAR imaging and also the necessity of a sufficient
CP insertion in an OFDM signal. In Section \ref{Conclusion}, we conclude this
paper and point out some future research problems.

\section{System Model and CP Based OFDM SAR Imaging}\label{System Model}
In this section, we first describe the OFDM SAR signal model, and then propose
the corresponding SAR imaging algorithm.

\subsection{OFDM SAR signal model}\label{Sub Sect1}
\begin{figure}[t]
\begin{center}
\includegraphics[width=0.7\columnwidth,draft=false]{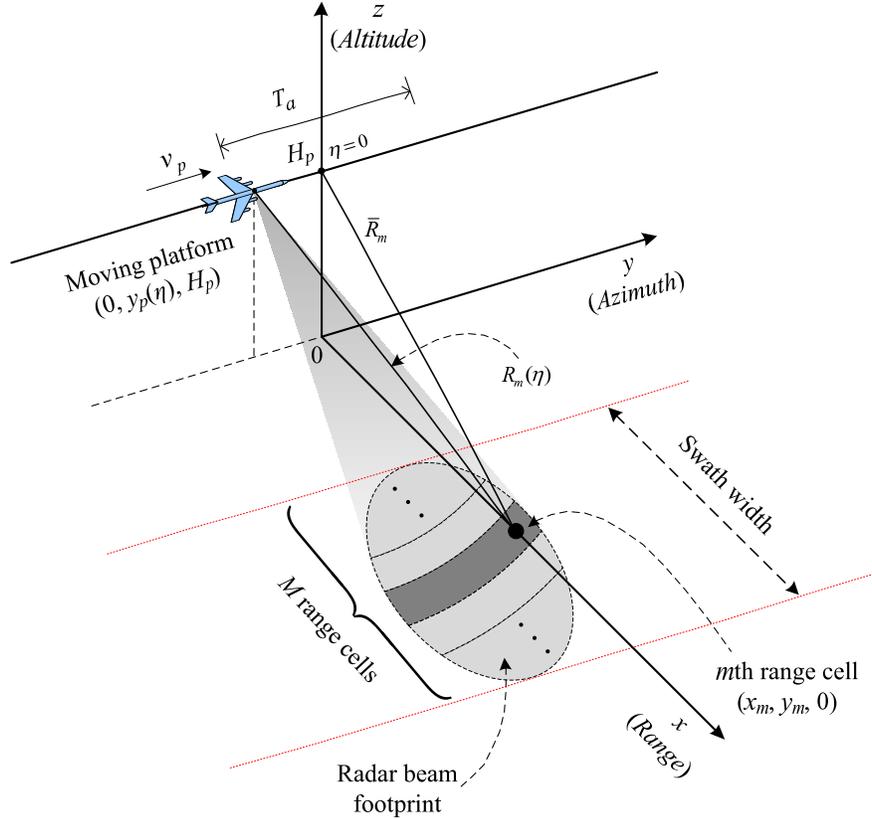}
\end{center}
\caption{Monostatic stripmap SAR geometry.}\label{geometry}
\end{figure}

In this paper, we consider the monostatic broadside stripmap SAR geometry as
shown in Fig. \ref{geometry}. The radar platform is moving parallelly to the
$y$-axis with an instantaneous coordinate $\left(0, y_p(\eta), H_p\right)$,
$H_p$ is the altitude of the radar platform, $\eta$ is the relative azimuth
time referenced to the time of zero Doppler, $T_a$ is the synthetic aperture
time defined by the azimuth time extent the target stays in the antenna beam.
For convenience, let us choose the azimuth time origin $\eta=0$ to be the zero
Doppler sample. Consider an OFDM signal with $N$ subcarriers, a bandwidth of
$B$ Hz, and let $\bS=\left[S_0,S_1,\ldots,S_{N-1}\right]^T$ represent the
complex weights transmitted over the subcarriers, and
$\sum\limits_{k=0}^{N-1}\left|S_k\right|^2=N$. Then, a discrete time OFDM
signal is the inverse fast Fourier transform (IFFT) of the vector $\bS$ and the
OFDM pulse is
\begin{equation}\label{OFDM} s\left(t\right)=\frac{1}{\sqrt{N}}\sum_{k=0}^{N-1}S_k
\textrm{exp}\left\{j2\pi k\Delta ft\right\},\ t\in \left[0,T+T_{GI}\right],
\end{equation}
where $\Delta f=\frac{B}{N}=\frac{1}{T}$ is the subcarrier spacing.
$\left[0,T_{GI}\right)$ is the time duration of the guard interval that
corresponds to the CP in the discrete time domain as we shall see later in more
details and its length $T_{GI}$ will be specified later too, $T$ is the length
of the OFDM signal excluding CP. Due to the periodicity of the exponential
function $\textrm{exp}(\cdot)$ in (\ref{OFDM}), the tail part of
$s\left(t\right)$ for $t$ in $\left(T,T+T_{GI}\right]$ is the same as the head
part of $s\left(t\right)$ for $t$ in $\left[0,T_{GI}\right)$. Let $f_c$ be the
carrier frequency of operation, the transmitted signal is given by
\begin{equation}
s_1\left(t\right)=\textrm{Re}\left\{\frac{1}{\sqrt{N}}\sum_{k=0}^{N-1}S_k\textrm{exp}\left\{j2\pi
f_kt\right\}\right\},\ t\in \left[0,T+T_{GI}\right],
\end{equation}
where $f_k=f_c+k\Delta f$ is the $k$th subcarrier frequency.

After the demodulation to baseband, the complex envelope of the received signal
from a static point target in the $m$th range cell can be written in terms of
fast time $t$ and slow time $\eta$
\begin{equation}\label{Receive}\begin{array}{ll}
u_m\left(t,\eta\right)&={g_m}\varepsilon_a(\eta)\textrm{exp} \left\{-j4\pi
f_c\frac{R_m(\eta)}{c}\right\}\\
&\times\frac{1}{\sqrt{N}}\sum\limits_{k=0}^{N-1}S_k
\textrm{exp}\left\{\frac{j2\pi k}{T}
\left[t-\frac{2R_m(\eta)}{c}\right]\right\}+w(t,\eta),\ t\in
\left[\frac{2R_m(\eta)}{c},\frac{2R_m(\eta)}{c}+T+T_{GI}\right],\\
\end{array}
\end{equation}
where $\varepsilon_a(\eta)=p_a^2\left(\theta(\eta)\right)$ is the azimuth
envelope,
$p_a(\theta)\approx\textrm{sinc}\left(\frac{0.886\theta}{\beta_{bw}}\right)$
\cite{Soumekh1999Synthetic}, $\textrm{sinc}(x)=\frac{\textrm{sin}(x)}{x}$ is
the sinc function, $\theta$ is the angle measured from boresight in the slant
range plane, $\beta_{bw}=\frac{0.866\lambda}{L_a}$ is the azimuth beamwidth,
$L_a$ is the effective length of the antenna, $g_m$ is the radar cross section
(RCS) coefficient caused from the scatterers in the $m$th range cell within the
radar beam footprint, and $c$ is the speed of light. $w(t,\eta)$ represents the
noise. $R_m(\eta)$ is the instantaneous slant range between the radar and the
$m$th range cell with the coordinate $(x_m, y_m, 0)$ and it can be written as
\begin{equation}
  R_m(\eta)=\sqrt{\bar{R}_m^2+\left(y_m-y_p\left(\eta\right)\right)^2}
  =\sqrt{\bar{R}_m^2+v_p^2\eta^2},
\end{equation}
where $\bar{R}_m=\sqrt{x_m^2+H_p^2}$ is the slant range when the radar platform
and the target in the $m$th range cell are the closest approach, and $v_p$ is
the effective velocity of the radar platform.

\begin{figure}[t]
\begin{center}
\includegraphics[width=0.6\columnwidth,draft=false]{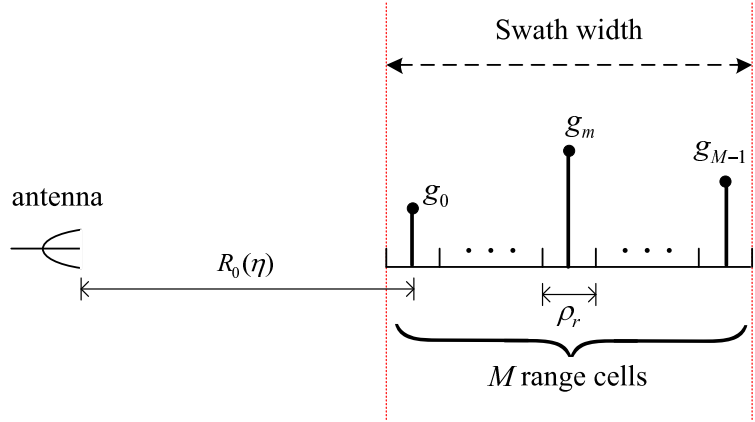}
\end{center}
\caption{Illustration diagram of one range line.}\label{RangeLine}
\end{figure}

Then, the complex envelope of the received signal from all the range cells in a
swath can be written as
\begin{equation}\label{ReceiveAll}
   u(t,\eta)=\sum_{m}u_m(t,\eta).
\end{equation}

At the receiver, with the A/D converter, the received signal is sampled with
sampling interval $T_s=\frac{1}{B}$ and the range resolution is
$\rho_r=\frac{c}{2B}$. Assume that the swath width for the radar is $R_w$. Let
$M=\frac{R_w}{\rho_r}$ that is determined by the radar system. Then, a range
profile can be divided into $M$ range cells as shown in Fig. \ref{RangeLine}.
As we mentioned earlier, the main reason why OFDM has been successfully adopted
in both recent wireline and wireless communications systems is its ability to
deal with multipaths (they cause ISI in communications) that become more severe
when the signal bandwidth is larger. In our radar applications here, the
response of each range cell, formed by the summation of the responses of all
scatterers within this range cell, contains its own delay and phase. Thus, to a
transmitted pulse, each range cell can be regarded as one path of
communications. $M$ range cells correspond to $M$ paths. Excluding one main
path (i.e., the nearest range cell), there will be $M-1$ multipaths. To convert
the ISI caused from the multipaths to the ISI free case in communications, a
guard interval (or CP) needs to be added to each OFDM block and the CP length
can not be smaller than the number of multipaths that is $M-1$ in this paper.
Although in the radar application here, ISI is not the concern, the $M$ range
cell paths are superposed (or interfered) together in the radar return signal,
which is the same as the ISI in communications. So, in order to convert these
interfered $M$ range cells to individual range cells without any IRCI, similar
to OFDM systems in communications, the CP length should be at least $M-1$. For
convenience, we use CP length $M-1$ in this paper, i.e., a CP of length $M-1$
is added at the beginning of an OFDM pulse, and then the guard interval length
$T_{GI}$  in the analog transmission signal is $T_{GI}=(M-1)T_s$. Notice that
$T=NT_s$, so the time duration of an OFDM pulse is $T_o=T+T_{GI}=(N+M-1)T_s$.
In this paper, we assume $N\geq M$, i.e., the number of subcarriers of the OFDM
signal is at least the number of range cells in a swath (or a cross range),
which is similar to the application in communications
\cite{prasad2004ofdm,goldsmith2005wireless}. When $N<M$, the IRCI occurs and
the detailed reason will be seen later.

\begin{figure}[b]
\begin{center}
\includegraphics[width=0.7\columnwidth,draft=false]{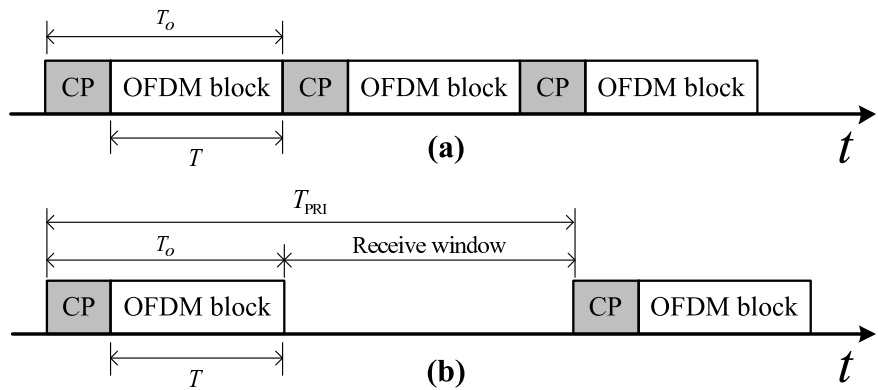}
\end{center}
\caption{Transmission comparison of OFDM signals in (a) communications systems
and (b) SAR systems.}\label{transmit_cycle}
\end{figure}

In communications applications, to achieve a high transmission throughput, the
OFDM pulses are transmitted consecutively as shown in Fig.
\ref{transmit_cycle}(a). However, in SAR imaging applications, for monostatic
case, the transmitter and receiver share the same antenna, which can not both
transmit and receive signals at the same time and transmission throughput is
not a concern. Thus, transmitted signals and radar return signals are usually
separated in time. This implies that a reasonable receive window is needed
between two consecutive pulses as shown in Fig. \ref{transmit_cycle}(b). For
convenience, similar to what is commonly done in SAR systems, in this paper, we
assume that the pulse repetition interval (PRI) is long enough so that all of
the range cells in a swath fall within the receive window. Therefore, the PRI
length $T_{\textrm{PRI}}$ should be
\begin{equation}
  T_{\textrm{PRI}}=\frac{1}{\textrm{PRF}}>\left(\frac{2R_w}{c}+T_o\right),
\end{equation}
where $R_w=M\rho_r$ is the swath width. We want to emphasize here that in our
common SAR imaging applications, the pulse repetition frequency (PRF) may not
be too high \cite{Soumekh1999Synthetic} and there is sufficient time duration
to add a CP (a guard interval) for an OFDM pulse.

For Fig. \ref{RangeLine}, we notice that $R_m(\eta)=R_0(\eta)+m\rho_r$. Thus,
$t-\frac{2R_m(\eta)}{c}$ in (\ref{Receive}) can be written as
\begin{equation}\label{TimeDelay}\begin{array}{ll}
  t-\frac{2R_m(\eta)}{c}&=t-\frac{2\left(R_0(\eta)+m\rho_r\right)}{c}\\
  &=t-t_0(\eta)-mT_s,\\
\end{array}\end{equation}
where for each $\eta$, the constant time delay $t_0(\eta)=\frac{2R_0(\eta)}{c}$
is independent of $m$. Let the sampling be aligned with the start of the
received pulse after $t_0(\eta)$ seconds for the first arriving version of the
transmitted pulse. Combining with (\ref{Receive}), (\ref{TimeDelay}) and
(\ref{ReceiveAll}), $u(t,\eta)$ can be converted to the discrete time linear
convolution of the transmitted sequence with the weighting RCS coefficients
$d_m$, and the received sequence can be written as
\begin{equation}\label{uig}
  u_i=\sum_{m=0}^{M-1}d_ms_{i-m}+w_i,\
  i=0,1,\ldots,N+2M-3,
\end{equation}
where
\begin{equation}\label{dm}
d_m=g_m\varepsilon_a(\eta)\textrm{exp} \left\{-j4\pi
f_c\frac{R_m(\eta)}{c}\right\},
\end{equation}
in which $4\pi f_c\frac{R_m(\eta)}{c}$ in the exponential is the azimuth phase,
and $s_i$ is the complex envelope of the OFDM pulse in (\ref{OFDM}) with time
duration $t\in \left[0,T+T_{GI}\right]$ for $T=NT_s$ and $T_{GI}=(M-1)T_s$.
After sampling at $t=iT_s$, (\ref{OFDM}) can be recast as:
\begin{equation}\label{OFDMn}
s_i=s\left(iT_s\right)=\frac{1}{\sqrt{N}}\sum_{k=0}^{N-1}S_k
\textrm{exp}\left\{\frac{j2\pi ki}{N}\right\},\ i=0,1,\ldots,N+M-2,
\end{equation}
and $s_i=0$ if $i<0$ or $i>N+M-2$.

Notice that the transmitted sequence with CP is
$\tilde{\bs}=\left[s_0,s_{1},\ldots,s_{N+M-2}\right]^T$, where
$\left[s_{0},\ldots,s_{M-2}\right]^T=\left[s_{N},\ldots,s_{N+M-2}\right]^T$.
The vector $\bs=\left[s_{0}, s_1,\ldots, s_{N-1}\right]^T$ is indeed the IFFT
of the vector $\bS=\left[S_0,S_1,\ldots,S_{N-1}\right]^T$.

\subsection{Range compression}\label{Sub Sect2}
When the signal in (\ref{uig}) is received, the first and the last $M-1$
samples\footnote{The reason to remove both the head and the tail $M-1$ samples
is because the total number of received signal samples in (\ref{uig}) is $N+
2(M-1)$. Because of the receive window between the OFDM pulses as shown in Fig.
\ref{transmit_cycle}(b), the tail $M-1$ samples in (\ref{uig}) are not affected
by the follow-up OFDM pulses. However, they do not have the full $M$ RCS
coefficients from all the $M$ range cells. If there is no receive window
between the OFDM pulses, the transmission is shown in Fig.
\ref{transmit_cycle}(a) as in communications, we only remove the head $M-1$
samples from the received signal sequence $u_n$ and use the next $N$ samples of
$u_n$ starting from $n=M-1$. } are removed, and then, we obtain
\begin{equation}\label{un}
u_n=\sum_{m=0}^{M-1}d_ms_{n-m}+w_n,\ n=M-1,M,\ldots,N+M-2.\\
\end{equation}

Then, the received signal $\bu=\left[u_{M-1},u_{M},\ldots,u_{N+M-2}\right]^T$
is
\begin{equation}\label{uHs_matrix}
  \begin{bmatrix}
  u_{M-1} \\u_{M}  \\ \vdots \\ u_{N+M-2}
  \end{bmatrix}
  =
  \begin{bmatrix}
  d_{M-1} & \cdots   & d_0     &\cdots &\cdots     &0       &\cdots  &0 \\
  0       & d_{M-1}  & \cdots  & d_0&\cdots &0   &\cdots  &0  \\
  \vdots  & \ddots   & \ddots  & \ddots & \ddots & \ddots & \ddots  &\vdots \\
  0       & \cdots   & 0  & \cdots   & \cdots     & d_{M-1}& \cdots & d_0
  \end{bmatrix}
  \begin{bmatrix}
  s_{0} \\ \vdots\\s_{M-1} \\s_{M} \\ \vdots \\ s_{N+M-2}
  \end{bmatrix}
  +
  \begin{bmatrix}
  w_{M-1} \\w_{M} \\ \vdots \\ w_{N+M-2}
  \end{bmatrix}.
\end{equation}
Since
$\left[s_{0},\ldots,s_{M-2}\right]^T=\left[s_{N},\ldots,s_{N+M-2}\right]^T$, it
is not hard to see that the vector $\tilde{\bs}=\left[s_0,s_{1},\ldots,\right.$
$\!\!\!\left. s_{N+M-2}\right]^T$ in (\ref{uHs_matrix}) can be replaced by its
tail part\footnote{This part is slightly different from what appears in
communications applications \cite[Ch. 5.2]{prasad2004ofdm}, \cite[Ch.
12.4]{goldsmith2005wireless} where the vector $\bs'$ in (\ref{resapeM}) is
replaced by the head part, $\bs$, of the vector $\tilde{\bs}$.} $\bs'$, which
can be also seen, in, for example, \cite[Ch. 12.4]{goldsmith2005wireless},
then, the matrix representation (\ref{uHs_matrix}) is equivalent to the
following representation:
\begin{equation}\label{resapeM}
  \bu  =\bH\bs'+\bw,
\end{equation}
where $\bs'=\left[s_{M-1},s_{M},\cdots,s_{N+M-2}\right]^T
=\left[s_{M-1},\cdots,s_{N-1},s_{0},\cdots,s_{M-2}\right]^T$,
$\bw=\left[w_{M-1},w_{M},\ldots,\right.$ $\!\!\!\left. w_{N+M-2}\right]^T$ and
$\bH$ is built by superposing the first $M-1$ columns of the weighting RCS
coefficient matrix in (\ref{uHs_matrix}) to its last $M-1$ columns. And $\bH$
can be given by the following $N$ by $N$ matrix:
\begin{equation}\label{Hmatrix}
  \bH=
  \begin{bmatrix}
  d_0     & 0      & \cdots & 0      & d_{M-1}& \cdots    & d_{1}     \\
  \vdots  & \ddots & \ddots & \vdots & \ddots & \ddots    & \vdots    \\
  d_{M-2} & \cdots & d_{0}  & 0      & \cdots & 0         & d_{M-1}   \\
  d_{M-1} &d_{M-2} & \cdots & d_{0}  & 0      & \cdots    & 0         \\
  0       & \ddots & \ddots & \vdots & \ddots & \ddots    & \vdots    \\
  \vdots  & \ddots & d_{M-1}& d_{M-2}& \cdots & d_{0}     & 0         \\
  0       & \cdots & 0      &d_{M-1} &d_{M-2} & \cdots    & d_0
  \end{bmatrix}.
\end{equation}
One can see that the matrix $\bH$ in (\ref{Hmatrix}) is a circulant matrix that
can be diagonalized by the discrete Fourier transform (DFT) matrix of the same
size.

The OFDM demodulator then performs a fast Fourier transform (FFT) on the vector
$\bu$:
\begin{equation}\label{U_k}
  U_k=\frac{1}{\sqrt{N}}\sum_{n=0}^{N-1}u_{n+M-1}\textrm{exp}
  \left\{\frac{-j2\pi kn}{N}\right\},\ k=0,1,\ldots,N-1.
\end{equation}
From (\ref{resapeM})-(\ref{Hmatrix}), the above $U_k$ can be expressed as:
\begin{equation}
  U_k=D_kS_k'+W_k,\ k=0,1,\ldots,N-1,
\end{equation}
where $\left[S_0',S_1',\cdots,S_{N-1}'\right]^T$ is the FFT of the vector
$\bs'$, a cyclic shift of the vector $\bs$ of amount $M-1$, i.e.,
$S_k'=S_k\textrm{exp}\left\{\frac{j2\pi k(M-1)}{N}\right\}$, $W_k$ is the FFT
of the noise, and
\begin{equation}\label{eDiM}
D_k=\sum_{m=0}^{M-1}d_{m}\textrm{exp}\left\{\frac{-j2\pi mk}{N}\right\}.
\end{equation}

Then, the estimate of $D_k$ is
\begin{equation}\label{hatD_i}
\hat{D}_k=\frac{U_k}{S_k'}=\frac{U_k}{S_k\textrm{exp}\left\{\frac{j2\pi
k(M-1)}{N}\right\}}=D_k+\frac{W_k}{S_k}\textrm{exp}\left\{\frac{-j2\pi
k(M-1)}{N}\right\},\ k=0,1,\ldots,N-1.
\end{equation}
Notice that if $S_k$ is small, the noise is enhanced. Thus, for the constraint
condition $\sum\limits_{k=0}^{N-1}\left|S_k\right|^2=N$, from (\ref{hatD_i})
the optimal signal $S_k$ should have constant module for all $k$.

The vector $\bD=\left[D_{0},D_1,\ldots,D_{N-1}\right]^T$ is indeed the
$N$-point FFT of $\sqrt{N} \bgamma$, where $\bgamma$ is the weighting RCS
coefficient vector:
\begin{equation}\label{dms}
\bgamma=\left[d_0,d_1,\cdots,d_{M-1},\underbrace{0,\cdots,0}_{N-M}\right]^T.
\end{equation}
So, the estimate of $d_m$ can be achieved by the $N$-point IFFT on the vector
$\hat{\bD}=\left[\hat{D}_{0},\hat{D}_1,\ldots,\hat{D}_{N-1}\right]^T$:
\begin{equation}\label{hatdtilde}
  \hat{d}_{m}=\frac{1}{\sqrt{N}}\sum_{k=0}^{N-1}\hat{D}_k\textrm{exp}
  \left\{\frac{j2\pi mk}{N}\right\},\ m=0,\ldots, M-1.\\
\end{equation}
Then, we obtain the following estimates of the $M$ range cell weighting RCS
coefficients:
\begin{equation}\label{estimated}
    \hat{d}_m={\sqrt{N}}d_m+\tilde{w}_m',\ m=0,\ldots, M-1,
\end{equation}
where $\tilde{w}_m'$ is from the noise and its variance is the same as that in
(\ref{hatD_i}) since the IFFT implemented in (\ref{hatdtilde}) is a unitary
transform. In $(\ref{estimated})$, $d_m$ can be completely recovered without
any IRCI from other range cells. From (\ref{dm}), when $d_m$ are determined,
the RCS coefficients $g_m$ are determined, and vice versa. 

\begin{figure}[t]
\begin{center}
\includegraphics[width=0.7\columnwidth,draft=false]{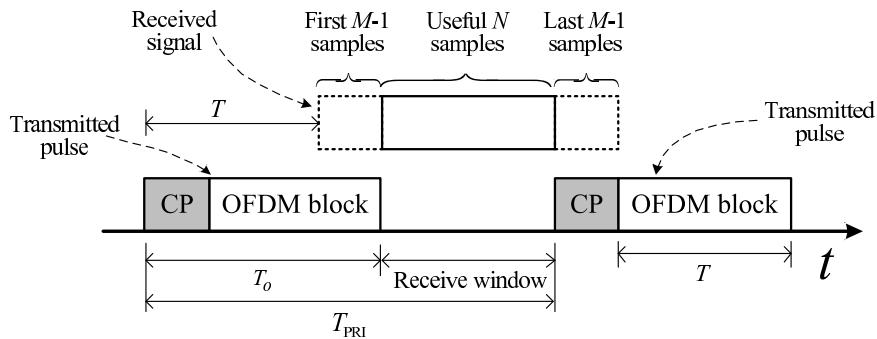}
\end{center}
\caption{Transmitted and received timing of OFDM SAR system for
$M=N$.}\label{received_cycle}
\end{figure}

For $N>M$, there are some zeros in the vector $\bgamma$ in (\ref{dms}).
Considering (\ref{dms}) and (\ref{hatdtilde}), we notice that part of the
transmitted OFDM sequence is used to estimate the unreal weighting RCS
coefficients, i.e., the zeros in $\bgamma$.

When $N=M$, there is no zeros included in the vector $\bgamma$ in (\ref{dms}).
We name this special case as {\em swath width matched pulse} (SWMP). The OFDM
pulse length (excluding CP) $T=NT_s$ and the swath width $M\rho_r$ follow the
relationship of $M\rho_r=\frac{cT}{2}$, i.e., the range resolution of a pulse
with time duration $T$ is just the swath width $M\rho_r$ without range
compression processing. And $\frac{cT}{2}$ is the maximal swath width that we
can obtain without IRCI. Thus, the optimal time duration of the OFDM pulse is
$T_o=(2N-1)T_s$ with CP length $N-1$, which is the maximal possible CP length
for an OFDM sequence of length $N$.

Since the first and the last $M-1$ samples of the received sequence $u_i$ in
(\ref{uig}) are removed, the receiver only needs to sample the received signal
from $t=T_0$ to $t=T_0+T$, even when the radar return signal starts to arrive
at time $t=T$ as shown in Fig. \ref{received_cycle}. Therefore, the minimum
range of the OFDM radar is $\frac{cT}{2}$, the same as that of the traditional
pulse radar of transmitted pulse length $T$. Notice that the minimum range is
just the same as the maximal swath width. Thus, if we want to increase the
swath width to, e.g., $10$ km, the transmitted pulse duration $T_o$ should be
increased to about $133.3\ \mu s$ and the minimum range is
increased\footnote{The pulse length here is much longer than the traditional
radar pulse and the number $N$ of subcarriers in OFDM signals is large too.
While long pulses of  LFM signals need  high frequency linearity and stability
and thus are not easy to generate, although long LFM pulses are not necessary,
long OFDM pulses (i.e., large $N$) are not difficult to generate since OFDM
signals can be easily generated by the IFFT operation as in (1). Moreover, the
conventional multiple channel SAR technology \cite{SuessAnovel20111013} for
wide imaged swath width can be used to reduce the OFDM pulse length and the
minimum range.} to $10$ km. Also, the minimum receive window is just the OFDM
pulse length $T$ (excluding CP) as Fig. \ref{received_cycle}, and the PRI
length follows $T_{\textrm{PRI}}>T_o+T$. As a remark, different from the
applications in communications where the CP is an overhead and may reduce the
transmission data rate, in the SWMP case in SAR applications here, the longer
the CP length is, the less the IRCI is, which leads to a better (high
resolution) SAR image.

When $N<M$, according to (\ref{OFDMn}) the signal vector
$\left[s_{0},\cdots,s_{M-1}, s_{M},\cdots,s_{N+M-2}\right]^T$ in
(\ref{uHs_matrix}) is $\left[s_{0},\ldots, s_{N-1},s_{0},\ldots,
s_{N-1},\ldots,s_{(N+M-2)_N}\right]^T$, where $(n)_N$ is the residue of $n$
modulo $N$. Thus, the $N$ by $N$ matrix $\bH$ in (\ref{resapeM}) and
(\ref{Hmatrix}) becomes
\begin{equation}
  \bH=
  \begin{bmatrix}
  \tilde{d}_0   & \tilde{d}_{N-1}  & \cdots  & \tilde{d}_1 \\
  \tilde{d}_1   & \tilde{d}_0      & \cdots  & \tilde{d}_2 \\
  \vdots        & \vdots           & \ddots  & \vdots      \\
  \tilde{d}_{N-1}& \tilde{d}_{N-2} & \cdots  & \tilde{d}_0
  \end{bmatrix},
\end{equation}
where $\tilde{d}_n=\sum\limits_{i:\ 0\leq iN+n\leq M-1}d_{iN+n},\
n=0,\ldots,N-1$. One can see, $\tilde{d}_n$ is a summation of weighting RCS
coefficients from several range cells, i.e., each $\tilde{d}_n$ has IRCI. Then,
following the OFDM approach (\ref{U_k})-(\ref{estimated}), what we can solve is
the superposed weighting RCS coefficients $\tilde{d}_n$, i.e., IRCI occurs.
This is the reason why we require $N\geq M$ in this paper.


After the range compression, combining the equations (\ref{Receive}),
(\ref{uig}), (\ref{dm}) and (\ref{estimated}), the range compressed signal can
be written as
\begin{equation}\label{Range compression}
u_{ra}(t,\eta)=\sqrt{N}\sum\limits_{m=0}^{M-1}\hat{g}_m\delta\left(t-\frac{2R_m(\eta)}{c}\right)
\varepsilon_a(\eta)\textrm{exp}\left\{-j4\pi
f_c\frac{R_m(\eta)}{c}\right\}+w_{ra}(t,\eta),
\end{equation}
where $\delta\left(t-\frac{2R_m(\eta)}{c}\right)$ is the delta function with
non-zero value at $t=\frac{2R_m(\eta)}{c}$, which indicates that the estimates
$\hat{g}_m$ of the RCS coefficient values ${g}_m$ are not affected by any IRCI
from other range cells after the range compression. $\hat{g}_m$ can be obtained
via (\ref{dm}) using the estimate $\hat{d}_m$ in (\ref{hatdtilde}). In the
delta function, the target range migration is incorporated via the azimuth
varying parameter $\frac{2R_m(\eta)}{c}$. Also, the azimuth phase in the
exponential is unaffected by the range compression.

Comparing with (\ref{Receive}), we notice that the range compression gain in
(\ref{Range compression}) is equal to $\sqrt{N}$, and the noise powers are the
same in (\ref{Range compression}) and (\ref{Receive}) when $S_k$ have constant
module. Thus, the signal-to-noise ratio (SNR) gain after the range compression
is $N$. For an LFM signal pulse with time duration $T_L=T_o=(N+M-1)T_s$ and the
same transmitted signal energy as in (\ref{OFDM}), it is well known that the
SNR gain after range compression is $N+M-1$, which is equal to the
time-bandwidth product (TBP) of the LFM signal pulse
\cite{Soumekh1999Synthetic}. Clearly, $N<N+M-1<2N$. This implies that the LFM
range compression SNR gain is larger (but not too much larger) than that of the
OFDM pulse. However, the IRCI exists because of the sidelobes of the ambiguity
function, resulting in a significant imaging performance degradation. In fact,
the sidelobe magnitude is roughly in the order of $\sqrt{N}$ in this case and
all the sidelobes from scatterers in all other $M-1$ range cells will be added
to the $m$th range cell for an arbitrary $m,\ 0\leq m\leq M-1$. One can see
that when $M$ is roughly more than $\sqrt{N}$, the scatterers in the $m$th
range cell will be possibly buried by the sidelobes of the scatterers in other
range cells and therefore can not be well detected and imaged. This is
similarly true for a random noise radar. In contrast, since the sidelobes are
ideally $0$ in the OFDM signal here, all $M$ scatterers can be ideally detected
and imaged without any IRCI as long as $N\geq M$, which may provide a high
range resolution image.

\subsection{Discussion on the design of weights $S_k$}\label{Discussion}
As one has seen from (\ref{U_k})-(\ref{estimated}), in order to estimate the
weighting RCS coefficients, the noise needs to be divided by $S_k$, which may
be significantly enhanced if $S_k$ is small. As mentioned earlier, in this
regard, the optimal weights $S_k$ should have constant module.

A special case of constant modular weights is that all $S_k$ are the same,
i.e., a constant. In this special case, the signal sequence $s_i$ to transmit
is the delta sequence, i.e., $s_0=\sqrt{N}$, and $s_i=0$ if $0<i\leq N-1$,
which is equivalent to the case of short rectangular pulse of pulse length
$\left[0,\ \frac{T}{N}\right]$. When a high range resolution is required, a
large bandwidth $B$ is needed and then there will be a large number $M$ of
range cells in a swath. This will require a large $N$. In this case, such a
short pulse with length $\left[0,\ \frac{T}{N}\right]$ and power $N$ may not be
easily implemented \cite{skolnik2001Introduction}. This implies that constant
weights $S_k$ may not be a good choice for the proposed OFDM signals.

Another case is when all the weights $S_k$ are completely random, i.e., they
are independently and identically distributed (i.i.d.). In this case, the mean
power of the transmitted signal $s_i$ is constant for every $i$. This gives us
the interesting property for an OFDM signal, namely, although its bandwidth is
as large as the short pulse of length $\left[0,\ \frac{T}{N}\right]$, its mean
energy is evenly spread over much longer ($N$ times longer) pulse duration,
which makes it much easier to generate and implement in a practical system than
the short pulse case.

In terms of the peak-to-average power ratio (PAPR) of transmitted signals, the
former case corresponds to the worst case, i.e., the highest PAPR case that is
$N$, while the later case corresponds to the best case in the mean sense, i.e.,
the lowest PAPR case that is $1$. After saying so, the above i.i.d. weight
$S_k$ case is only in the statistical sense. In practice, a deterministic
weight sequence $S_k$ is used, which can be only a pseudo-random noise (PN)
sequence and therefore, its $N$-point inverse discrete Fourier transform $s_i$
(and/or its analog waveform $s(t)$) may not have a constant power and in fact,
its PAPR may be high (although may not be the highest) compared with the LFM
radar or the random noise radar. This will be an interesting future research
problem on how to deal with the high PAPR problem of OFDM signals for radar
applications. Note that there have been many studies for the PAPR reduction in
communications community, see for example
\cite{prasad2004ofdm,goldsmith2005wireless}. If we only consider the finite
time domain signal values, i.e., the IDFT, $s_i$, of the weights $S_k$ in
(\ref{OFDMn}), we can use a Zadoff-Chu sequence as $S_k$ that is, in fact, a
discrete LFM signal, and then its IDFT, $s_i$, has constant module as well
\cite{BeymeEfficientComputationEL2009461, PopovicEfficientDFTEL2010502}. In
this case, both the weights $S_k$ and the discrete time domain signal values
$s_i$ have constant module, i.e., the discrete PAPR (the peak power over the
mean power of $s_i$) is $1$.

As a remark, if one only considers the discrete transmitted signal sequence
$s_i$, it can be, in fact, from any radar signal, such as LFM or random noise
radar signal as follows.

Let $s'\left(t\right)$ be any radar transmitted signal and set
$s_i=s'\left(\frac{iT}{N}\right),\ 0\leq i\leq N-1$. Then, we can always find
the corresponding weights $S_k,\ 0\leq k\leq N-1$, that are just the $N$-point
FFT of $s_i,\ 0\leq i\leq N-1$. Then, the analog OFDM waveform
$s\left(t\right)$ in (\ref{OFDM}) can be thought of  as an interpolation of
this discrete time sequence $s_i$. Thus, as $N$ goes large, the analog OFDM
waveform $s\left(t\right)$ can approach the given radar waveform
$s'\left(t\right)$.

\subsection{Insufficient CP case}
Let us consider the CP length to be $\bar{M}$ and $\bar{M}<M-1$. In this case,
the length of CP is insufficient. When an insufficient CP is used, if OFDM
pulses are transmitted consecutively without any waiting interval as in
communications systems, the OFDM blocks will interfer each other due to the
multipaths, which is called inter-block-interference (IBI). However, this will
not occur in our radar application in this paper, since the second OFDM pulse
needs to wait for receiving all the radar return signals of the first
transmitted OFDM pulse as we have explained in Section \ref{System Model}.A
earlier. Although there is no IBI, the insufficient CP will cause the
inter-carrier-interference (ICI) that leads to the IRCI as shown below. In this
case, (\ref{resapeM}) can be recast as
\begin{equation}\label{resapeMSu}
\begin{array}{ll}
  \Delta\bu&=\bu-\bar{\bu}\\
  &=\bH\bs'-\bar{\bH}\bs'+\bw,
\end{array}
\end{equation} where
\begin{equation}
\bar{\bH}=
  \begin{bmatrix}
  0      & \cdots & d_{M-1}& d_{M-2}& \cdots & d_{\bar{M}+1} &0 & \cdots &0       \\
  0      & \cdots & 0      & d_{M-1}& \cdots & d_{\bar{M}+2} &0 & \cdots &0       \\
  \vdots & \cdots & \vdots & \ddots & \ddots & \vdots  & \vdots & \vdots & \vdots \\
  0      & \cdots & 0      & 0      & \ddots & d_{M-1} & 0      & \cdots &0       \\
  \vdots & \cdots & \vdots & \ddots & \ddots & \vdots  & \vdots & \cdots & \vdots \\
  0      & \cdots & 0      & 0      & \cdots & 0       & 0      & \cdots &0       \\
  \end{bmatrix}.
\end{equation}

The $n$th element of $\Delta\bu$ can be expressed as
\begin{equation}
\Delta u_n=u_{n}-\bar{u}_{n}, \ n=M-1,M,\ldots,N+M-2,
\end{equation}
where $u_n$ is the same as (\ref{un}), i.e., from $\bH \bs'$ and noise $\bw$,
and
\begin{equation}\label{bar_un}
\bar{u}_{n}=\sum_{\tilde{m}=\bar{M}-M+2+n}^{M-1}d_{\tilde{m}}s_{n-\tilde{m}}.\\
\end{equation}
Notice that $\bar{u}_{n}=0$ for $n=2M-\bar{M}-2,\cdots,N+M-2$. Then, the
$N$-point FFT is performed on the vector $\Delta\bu$
\begin{equation}
  U_k=D_kS_k-\bar{U}_k+W_k,\ k=0,1,\ldots,N-1,
\end{equation}
where
\begin{equation}
\bar{U}_k=\frac{1}{\sqrt{N}}\sum_{n=0}^{N-1}\bar{u}_{n+M-1}\textrm{exp}\left\{\frac{-j2\pi
kn}{N}\right\}.
\end{equation}

After the $N$-point IFFT is performed on the vector
$\left[\frac{\bar{U}_0}{S_0'},\frac{\bar{U}_1}{S_1'},\cdots,\frac{\bar{U}_{N-1}}{S_{N-1}'}\right]^T$,
we can obtain
\begin{equation}
 \hat{d}_m=\sqrt{N}d_m-\xi_m+\tilde{w}_m,\ m=0,1,\ldots,M-1,
\end{equation}
where
\begin{equation}\label{xi_m}
\xi_m=\frac{1}{N}\sum_{k=0}^{N-1}\frac{1}{S_k}\sum_{n=0}^{M-\bar{M}-2}
\sum_{\tilde{m}=\bar{M}-M+2+n}^{M-1}d_{\tilde{m}}s_{n+M-1-\tilde{m}}\textrm{exp}
\left\{\frac{-j2\pi k(n-m)}{N}\right\}.
\end{equation}

We remark that $\xi_m$ is the IRCI resulted from the insufficient CP, and
$\xi_m$ is related to the reflectivities of the neighboring range cells and the
transmitted signal. From the second and third summation signs in $\xi_m$ in
(\ref{xi_m}) and the equation (\ref{bar_un}), a smaller $\bar{M}$ leads to more
range cells involved in the interference, resulting in a stronger IRCI. The
performance degradation with different insufficient CP lengths of $\bar{M}$
will be shown in the simulations in Section \ref{Simulations} later.

\section{Simulations and Performance Discussions}\label{Simulations}
\begin{figure}[t]
\begin{center}
\includegraphics[width=0.6\columnwidth,draft=false]{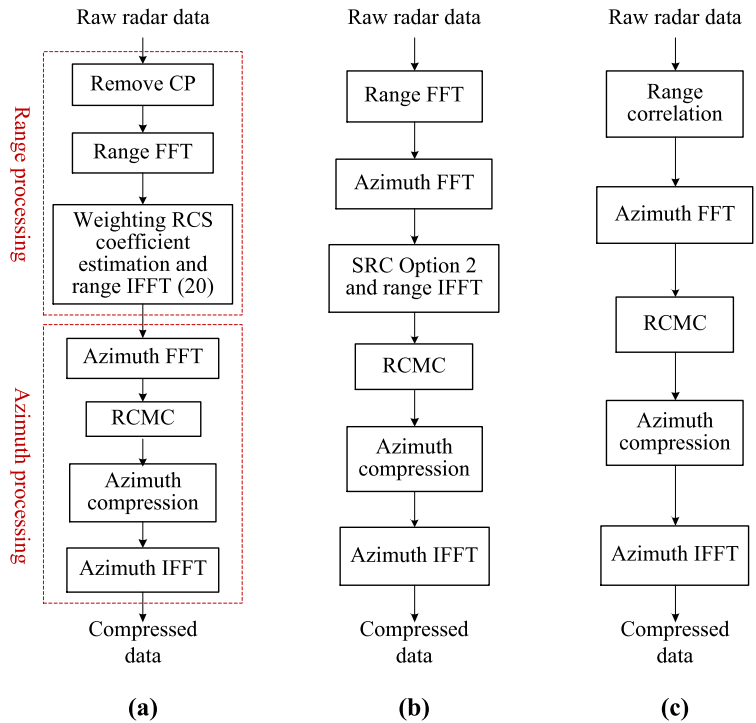}
\end{center}
\caption{Block diagram of SAR imaging processing: (a) CP based OFDM SAR; (b)
LFM SAR; (c) Random noise SAR and conventional OFDM SAR.}\label{LFM_Block}
\end{figure}

This section is to present some simulations and discussions for our proposed CP
based OFDM SAR imaging. The simulation stripmap SAR geometry is shown in Fig.
\ref{geometry}. The azimuth processing is similar to the conventional stripmap
SAR imaging \cite{Soumekh1999Synthetic} as shown in Fig. \ref{LFM_Block}(a).
For computational efficiency, a fixed value of $R_c$ located at the center of
the range swath is set as the reference range cell as
\cite{Soumekh1999Synthetic}. Then, the range cell migration correction (RCMC)
and the azimuth compression are implemented in the whole range swath using
$R_c$. For convenience, we do not consider the noise in this section as what is
commonly done in SAR image simulations. For comparison, we also consider the
range Doppler algorithm (RDA) using LFM and random noise signals as shown in
the block diagram of Fig. \ref{LFM_Block}. Since the performance of a step
frequency signal SAR is similar to that of an LFM signal SAR, here we only
consider LFM signal SAR in our comparisons. In Fig. \ref{LFM_Block} (b), the
secondary range compression (SRC) is implemented in the range and azimuth
frequency domain, the same as the Option 2 in \cite[Ch.
6.2]{Soumekh1999Synthetic}. In Fig. \ref{LFM_Block} (c), the range compression
of the random noise signal and the conventional OFDM signal are achieved by the
correlation between the transmitted signals and the range time domain data.
Notice that the difference of these three imaging methods in Fig.
\ref{LFM_Block} is the range compression, while the RCMC and azimuth
compression are identical.

The simulation experiments are performed with the following parameters as a
typical SAR system: PRF = $800$ Hz, the bandwidth is $B = 150$ MHz, the antenna
length is $L_a=1$ m, the carrier frequency $f_c=9$ GHz, the synthetic aperture
time is $T_a=1$ sec, the effective radar platform velocity is $v_p=150$ m/sec,
the platform height of the antenna is $H_p=5$ km, the slant range swath center
is $R_c=5\sqrt{2}$ km, the sampling frequency $f_s=150$ MHz, the number of
range cells is $M=96$ with the center at $R_c$. For the convenience of FFT/IFFT
computation, we set $T=\frac{512}{150}\ \mu s\approx 3.41\ \mu s$, then the
number of subcarriers for the OFDM signal is $N=512$. The CP length is $95$
that is sufficient and the CP time duration is $T_{GI}=\frac{95}{150}\ \mu
s\approx 0.63\ \mu s$. Thus, the time duration of an OFDM pulse is
$T_o=\frac{607}{150}\ \mu s\approx 4.05\ \mu s$. The complex weight vectors
over the subcarriers of the CP based OFDM signal and the conventional OFDM
signal are set to be vectors of binary PN sequence of values $-1$ and $1$.
Meanwhile, for the transmission energies of the three SAR imaging methods to be
the same, the time durations of an LFM pulse, a conventional OFDM pulse and a
random noise pulse are also $4.05\ \mu s$.

\begin{figure}[t]
\begin{center}
\includegraphics[width=0.7\columnwidth,draft=false]{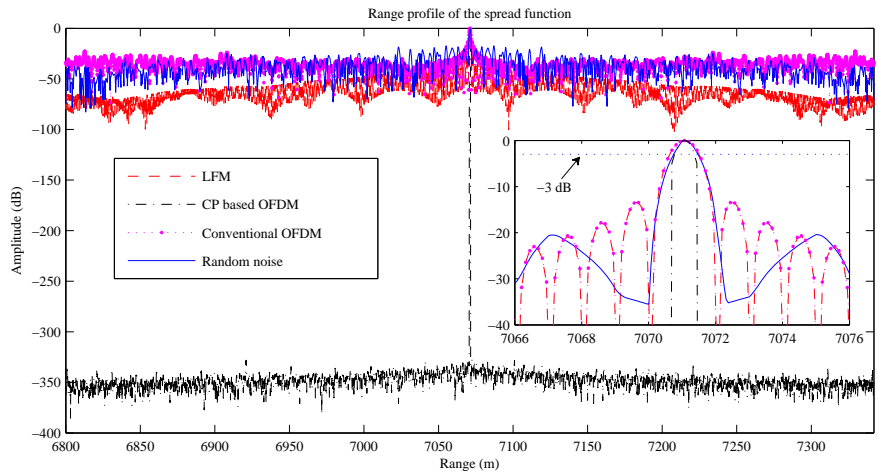}
\end{center}
\caption{The range profiles of the point spread function.}\label{range_profile}
\end{figure}

A point target is assumed to be located at the range swath center. Without
considering the additive noise, the normalized range profiles of the point
spread function are shown in Fig. \ref{range_profile} and the details around
the mainlobe area are shown in its zoom-in image. It can be seen that the
sidelobes are much lower for the CP based OFDM signal than those of the other
three signals, while the $3$ dB mainlobe widths of the four signals are all the
same.

\begin{figure}[t]
\begin{center}
\includegraphics[width=0.7\columnwidth,draft=false]{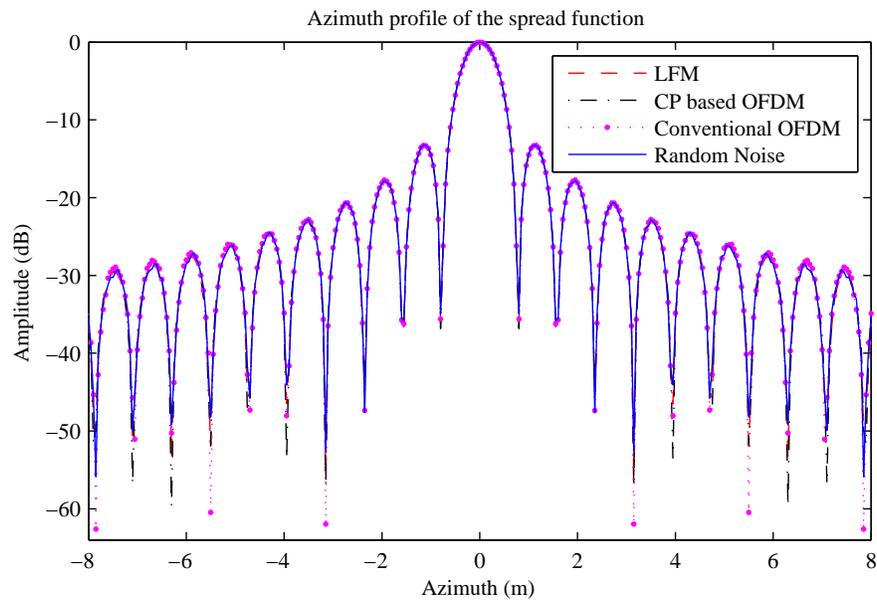}
\end{center}
\caption{The azimuth profiles of the point spread function.}\label{azimuth
profile}
\end{figure}

The normalized azimuth profiles of the point spread function of the three
methods are shown in Fig. \ref{azimuth profile}. The results show that the
azimuth profiles of the point spread function are similar for all the four
signals of LFM, CP based OFDM, conventional OFDM and random noise.

\begin{figure}[b]%
\centering \vspace{-0.2in}\subfigure[]{
\label{Tank:a} 
\includegraphics[width=2.3in]{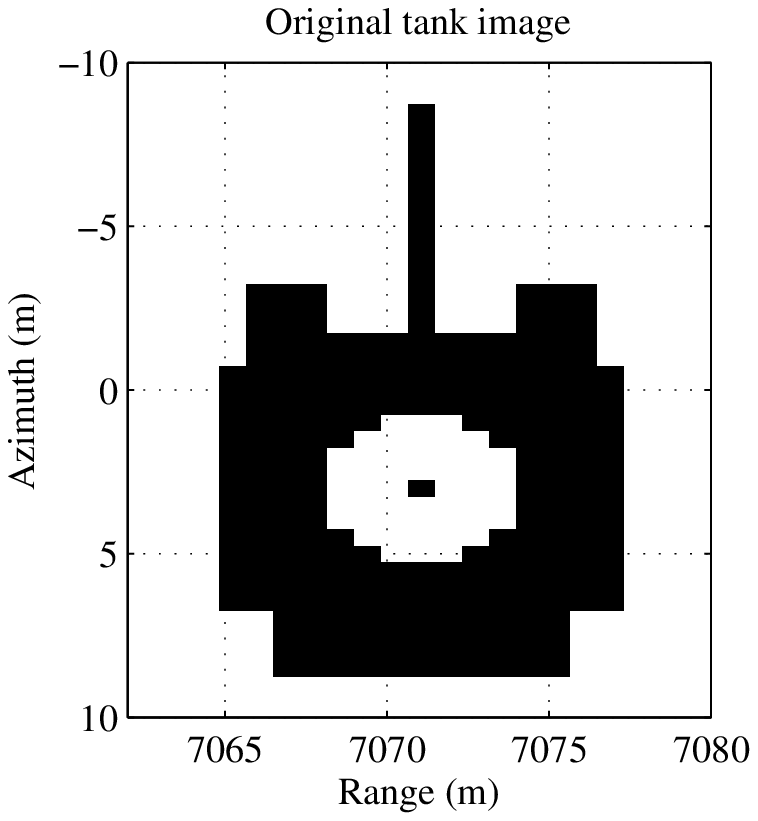}}
 \hspace{0.07in} \subfigure[]{
\label{Tank:b} 
\includegraphics[width=2.3in]{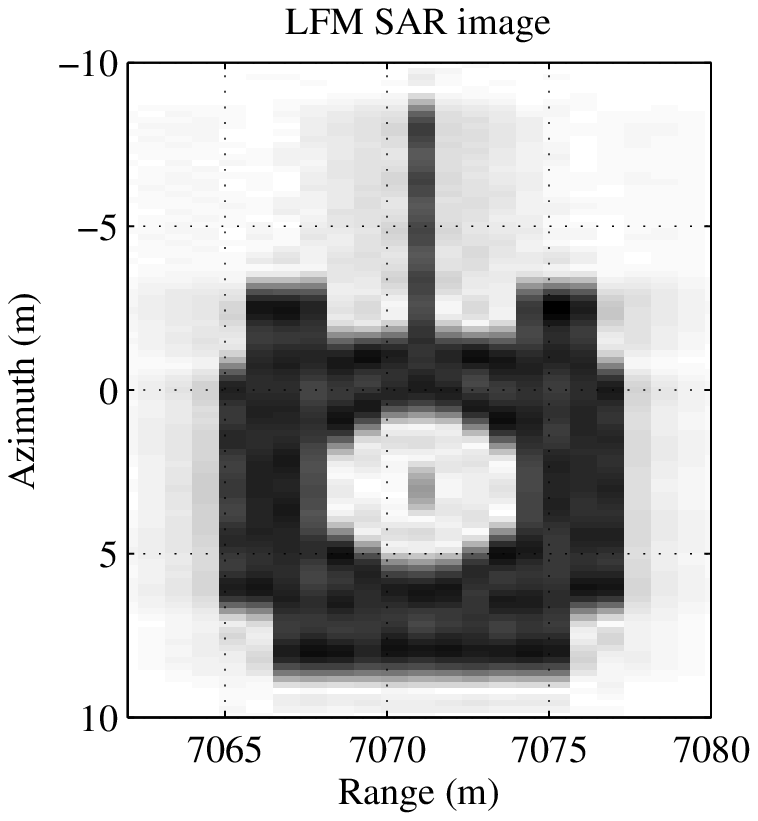}}\\
\subfigure[]{
\label{Tank:c} 
\includegraphics[width=2.3in]{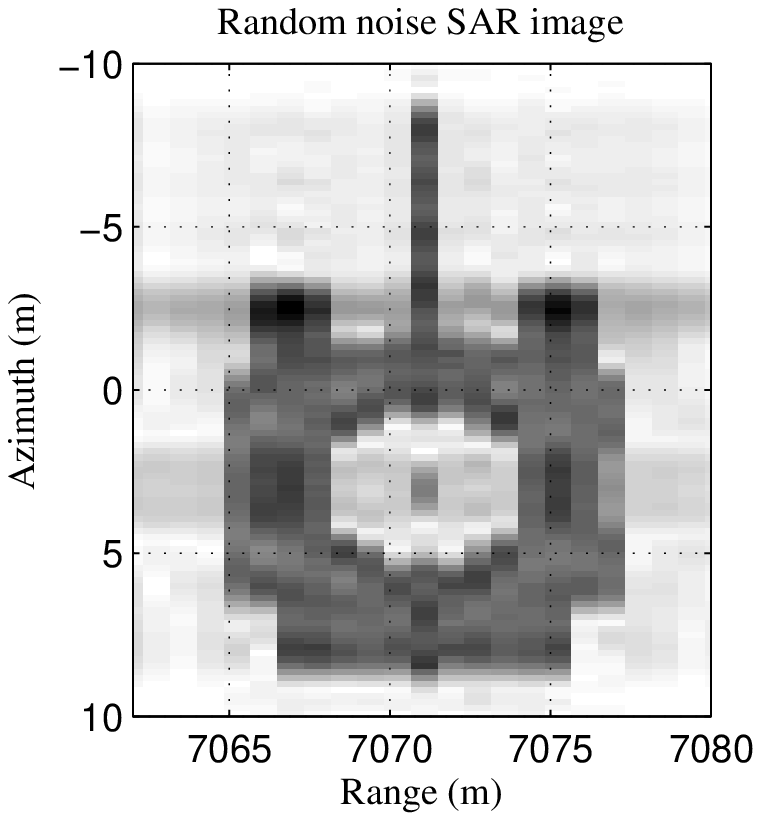}}
\hspace{0.07in} \subfigure[]{
\label{Tank:d} 
\includegraphics[width=2.3in]{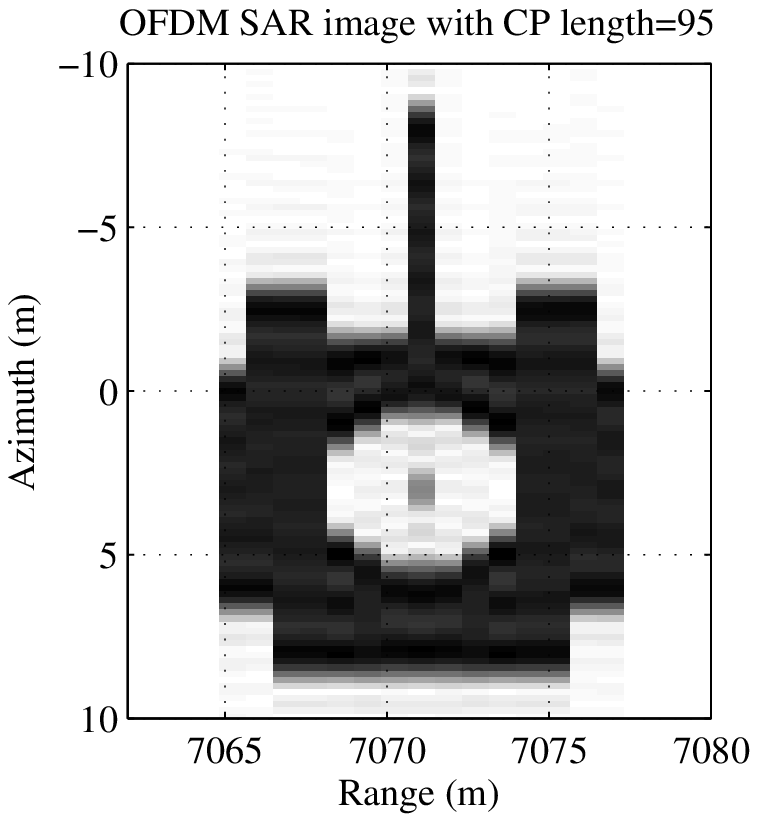}}\\
\label{Tank:e} 
\subfigure[]{
\includegraphics[width=2.3in]{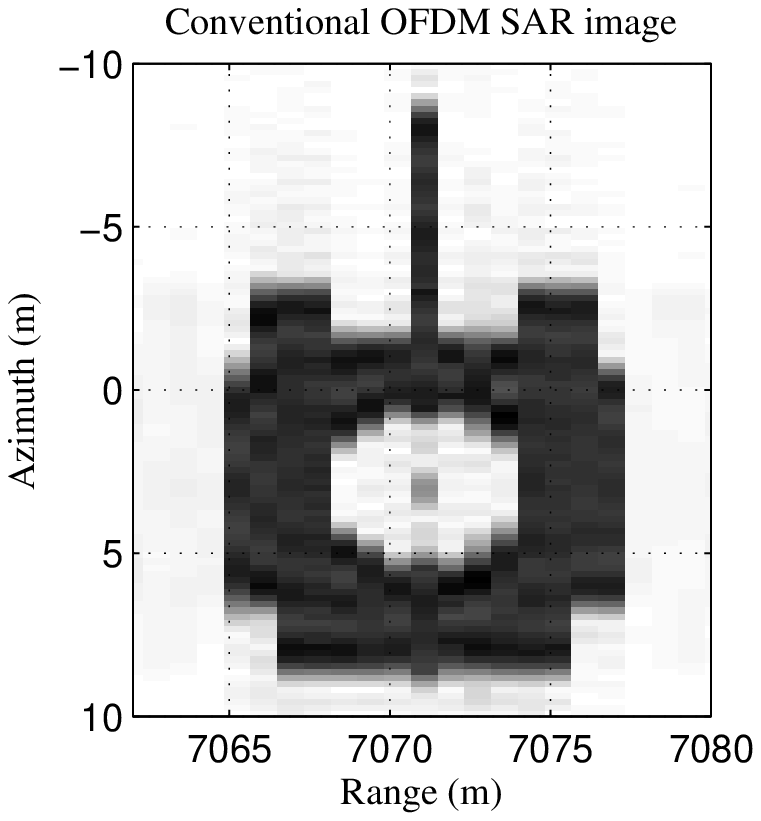}}
\hspace{0.07in} \subfigure[]{
\label{Tank:f} 
\includegraphics[width=2.3in]{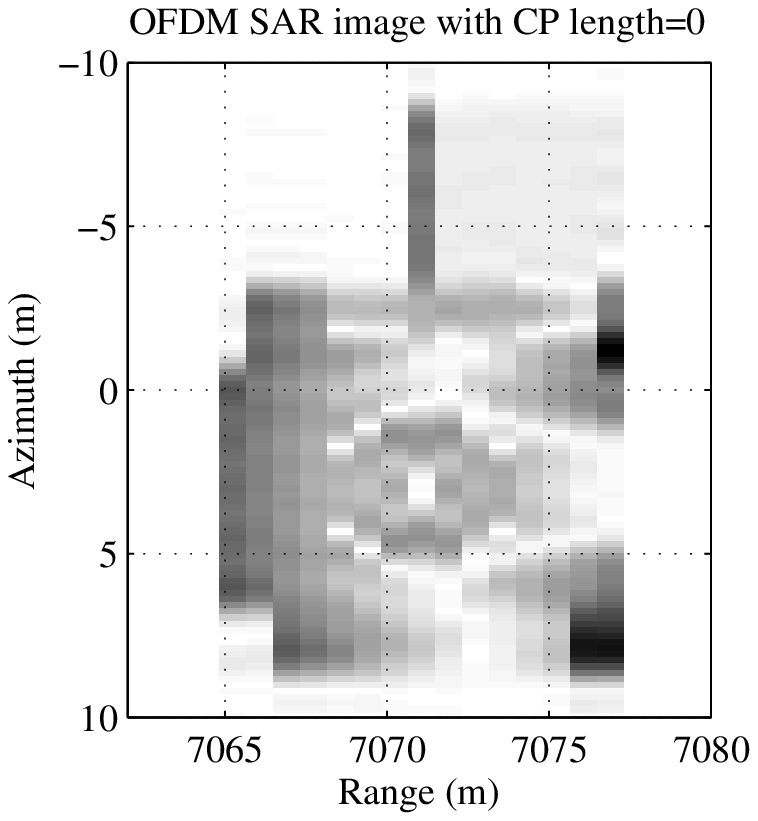}}
\caption{Imaging results of simulated reflectivity profile for a tank: (a)
Original tank; (b) LFM SAR; (c) Random noise SAR; (d) OFDM SAR with sufficient
CP; (e) Conventional OFDM signal SAR; (f) CP based OFDM SAR with CP length
$=0$.} \label{Tank}
\end{figure}

We then consider an extended object with the shape of a tank constructed by a
few single point scatterers, and the original reflectivity profile is shown as
Fig. \ref{Tank:a}. The results indicate that the imaging performance using the
CP based OFDM SAR is better than the other three signals. Specifically, the
boundaries of the extended object are observed less blurred by using the CP
based OFDM SAR imaging. In Fig. \ref{Tank:f}, we also consider the imaging of
the tank with our proposed method when the CP length is zero and the
transmitted OFDM pulse is the same as the conventional OFDM pulse. As a remark,
comparing Fig. \ref{Tank}(e) and Fig. \ref{Tank}(f), one may see that the SAR
imaging performance degradation is significant when CP length is zero. This is
because our proposed range reconstruction method in the receiver, as mentioned
at (\ref{U_k})-(\ref{estimated}), is for CP based OFDM SAR imaging and
different with the traditional matched filter SAR imaging method. Thus,
sufficient CP should be included in the transmitted OFDM pulse to achieve IRCI
free range reconstruction.

\begin{figure}[t]
\begin{center}
\includegraphics[width=0.9\columnwidth,draft=false]{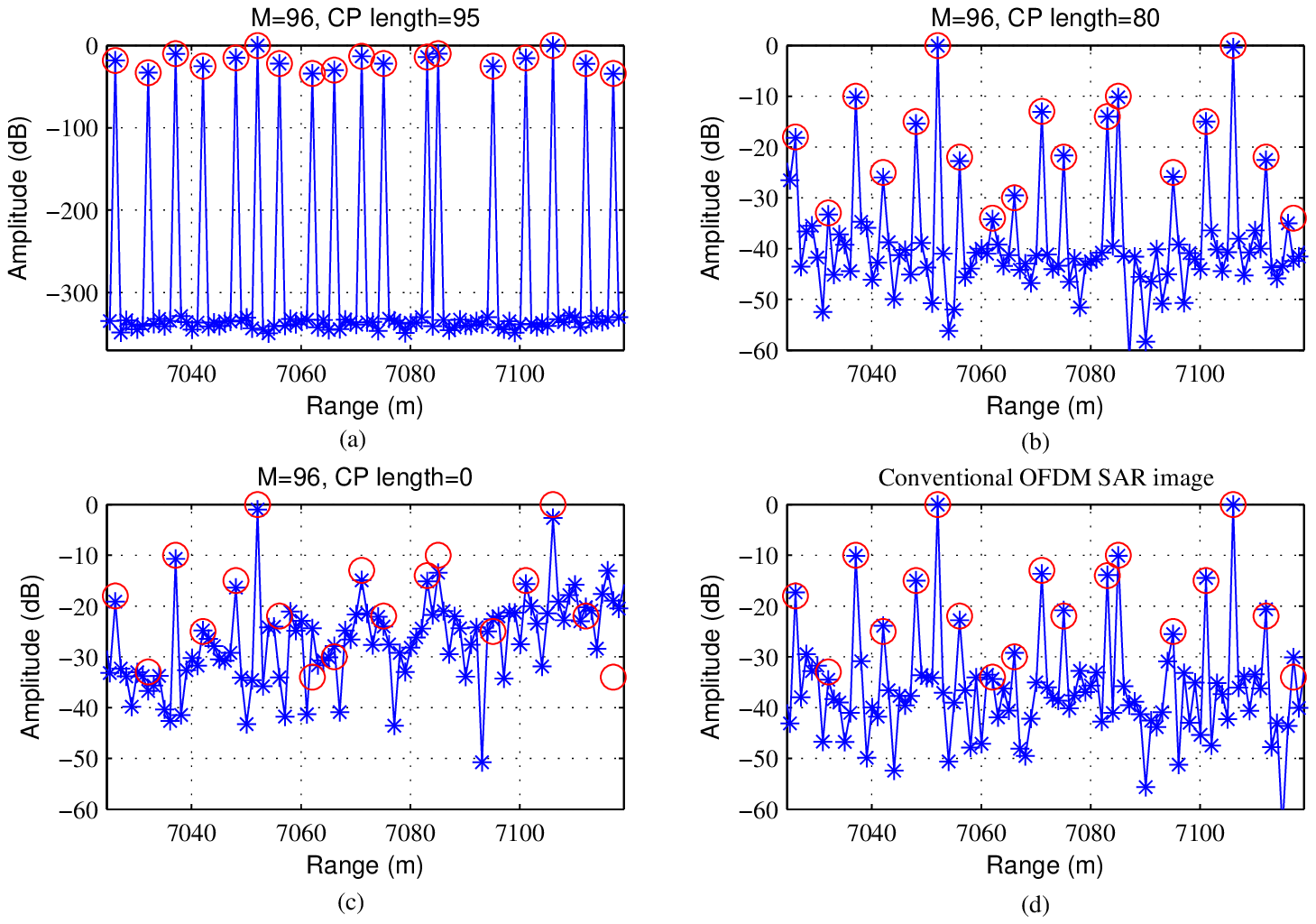}
\end{center}
\caption{A range line imaging with different CP lengths. Red circles denote the
real target amplitudes, blue asterisks denote imaging
results.}\label{RangeLine_InCP}
\end{figure}

We next consider the importance of adding a sufficient CP in our proposed CP
based OFDM SAR imaging. We consider a single range line (a cross range) with
$M=96$ range cells, and targets are included in $18$ range cells, the
amplitudes are randomly generated and shown as the red circles in Fig.
\ref{RangeLine_InCP}, and the RCS coefficients of the other range cells are set
to be zero. The normalized imaging results are shown as the blue asterisks. The
results indicate that the imaging is precise when the length of CP is $95$,
i.e., sufficient CP length, in Fig. \ref{RangeLine_InCP}(a), the amplitudes of
the range cells without targets are lower than $-300$ dB, which are due to the
computer numerical errors. With the decrease of the CP length the imaging
performance is degraded and the IRCI is increased. Specifically, the zero
amplitude range cells become non-zero anymore and some targets are even
submerged by the IRCI as shown in Fig. \ref{RangeLine_InCP}(b) and Fig.
\ref{RangeLine_InCP}(c). We also show the imaging results with the conventional
OFDM SAR image as in Fig. \ref{RangeLine_InCP}(d). The curves in Fig.
\ref{RangeLine_InCP}(d) indicate that some targets are submerged by the IRCI
from other range cells. In Fig. \ref{RangeLine_InCP}, we notice that, when the
CP lengthes are $95$ and $80$ (as in Fig. \ref{RangeLine_InCP}(a) and Fig.
\ref{RangeLine_InCP}(b), respectively), the imaging performances of our
proposed method outperform the conventional OFDM SAR image in Fig.
\ref{RangeLine_InCP}(d). However, the imaging performance with zero length CP
is worse than the conventional OFDM SAR image, although they have the same
transmitted OFDM waveform, which is again because our proposed range
reconstruction method at the receiver is different from the conventional
method. It also further indicates that a sufficient CP is important for our
proposed CP based OFDM SAR imaging.

\begin{figure}[t]
\begin{center}
\includegraphics[width=0.7\columnwidth,draft=false]{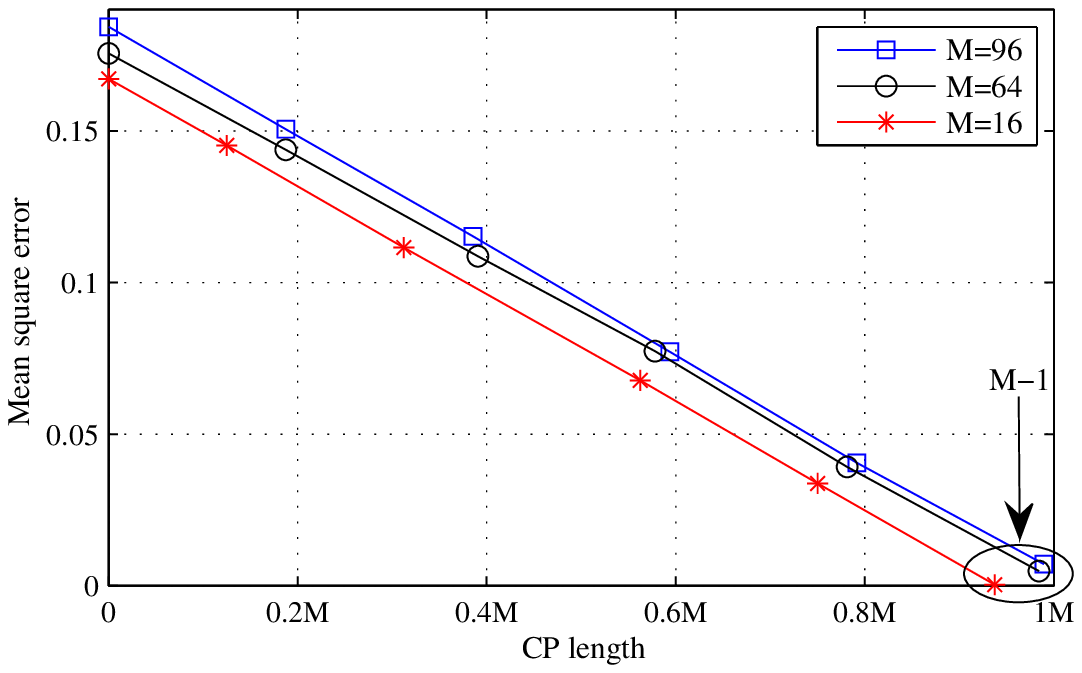}
\end{center}
\caption{The mean square errors for insufficient CP lengths.}\label{LessCP}
\end{figure}

We also consider a single range line (a cross range) with $M$ range cells, in
which the RCS coefficients are set $g_m=1,\ m=0,\ldots,M-1$. After the CP based
OFDM SAR imaging with different lengths of CP, we calculate the mean square
errors (MSE) between the energy normalized imaging results and the original RCS
coefficients $g_m$. The results are achieved from the average of $1000$
independent Monte Carlo simulations and are shown in Fig. \ref{LessCP}. The
curves suggest that the performance degradation occurs when the length of CP is
less than $M-1$, i.e., insufficient. The MSE is supposed to be zero when the CP
length is $M-1$ that is sufficient. However, one can still observe some errors
in Fig. \ref{LessCP}, which is because errors may occur by using a fixed
reference range cell $R_c$ in the imaging processing (i.e., RCMC and azimuth
compression), and the wider swath (or a larger $M$) causes the larger imaging
error. Thus, the MSE is slightly larger when $M$ is larger.

\section{Conclusion}\label{Conclusion}

In this paper, by using the most important feature of OFDM signals in
communications systems, namely converting an ISI channel to multiple ISI-free
subchannels, we have proposed a novel method for SAR imaging using OFDM signals
with sufficient CP. The sufficient CP insertion provides an IRCI free (high
range resolution) SAR image. We first established the CP based OFDM SAR imaging
system model and then derived the CP based OFDM SAR imaging algorithm with
sufficient CP and showed that this algorithm has zero IRCI (or IRCI-free) for
each cross range. We also analyzed the influence when the CP length is
insufficient. By comparing with the LFM SAR and the random noise SAR imaging
methods, we then finally provided some simulations to illustrate the high range
resolution property of the proposed CP based OFDM SAR imaging and also the
necessity of a sufficient CP insertion in an OFDM signal. The main features of
the proposed SAR imaging are highlighted below.
\begin{itemize}
  \item
  The sufficient CP length $M-1$ is determined by the number of range cells $M$ within
  a swath, which is directly related to the range resolution of the SAR system.
  \item
  The optimal time duration of the OFDM pulse is $T_o=(2N-1)T_s$ with CP length $N-1$.
  The minimum range of the proposed CP based OFDM SAR is the same as the maximal swath
  width.
  \item
  The range sidelobes are ideally zero for the proposed CP based OFDM SAR imaging, which
  can provide high range resolution potential for SAR systems. From our simulations, we
  see that the imaging performance of the CP based OFDM SAR is better than those of the
  LFM SAR and the random noise SAR, which may be more significant in MIMO radar applications.
  \item
  The imaging performance of the CP based OFDM SAR is degraded and the IRCI is increased
  when the CP length is insufficient.
\end{itemize}

Some future researches may be needed for our proposed CP based OFDM SAR imaging
systems. One of them is the high PAPR problem of the OFDM signals as what has
been pointed out in Section \ref{Discussion}. Another problem is that OFDM
signals are sensitive to an unknown Doppler shift, such as, that induced from
an known moving target, which may damage the orthogonality of OFDM subcarriers
and result in the  potential ICI. Also, how does our proposed OFDM SAR imaging
method work for a distributed target model
\cite{KriegerMIMOSAROpportunitiesIEEETGRS20142628}? Some of these problems are
under our current investigations.

\section*{Acknowledgment}
The authors would like to thank the editor and the anonymous reviewers for
their useful comments and suggestions that have improved the presentation of
this paper.

\clearpage
\bibliographystyle{IEEEtran}
\bibliography{IEEEabrv,D:/assignment/Report/Paper/MyReference}
\end{document}